\documentclass[prd,aps,amssymb,amsmath,tightenlines,showpacs]{revtex4}

\usepackage{graphicx,epsfig}

\def\half{{\textstyle{\frac{1}{2}}}}
\def\cP{\mathcal P}

\def\cT{\mathcal T}
\def\cPT{\mathcal PT}
\def\w{\omega}

\begin{document}

\title{Time-independent Hamiltonian for any linear constant-coefficient
evolution equation}

\author{Carl~M.~Bender$^a$}\email{cmb@wustl.edu}
\author{Mariagiovanna~Gianfreda$^b$}\email{gianfred@le.infn.it}
\author{Nima~Hassanpour$^a$}\email{nimahassanpourghady@wustl.edu}
\author{Hugh~F.~Jones$^c$}\email{h.f.jones@imperial.ac.uk}

\affiliation{$^a$Department of Physics, Washington University, St. Louis, MO
63130, USA\\
$^b$Institute of Industrial Science, University of Tokyo, Komaba, Meguro,
Tokyo 153-8505, Japan\\
$^c$Theoretical Physics, Imperial College London, London SW7 2AZ, UK}

\begin{abstract}
It is shown how to construct a time-independent Hamiltonian having only one
degree of freedom from which an arbitrary linear constant-coefficient evolution
equation of any order can be derived.
\end{abstract}

\date{\today}
\maketitle

\section{introduction}\label{s1}
If one were to ask a randomly chosen physicist whether the equation of motion of
the damped classical harmonic oscillator
\begin{equation}
\ddot{x}+2\gamma\dot{x}+\w^2x=0\qquad(\gamma>0)
\label{e1}
\end{equation}
can be derived from a {\it time-independent} Hamiltonian, almost certainly the
answer would be a resounding ``no,'' because multiplying this equation by
$\dot{x}$, one obtains
the equation
\begin{equation}
\frac{d}{dt}\left(\half\dot{x}^2+\half\w^2x^2\right)=-2\gamma\dot{x}^2.
\label{e2}
\end{equation}
The quantity $\half\dot{x}^2+\half\w^2x^2$, which appears to be the sum of a
kinetic and a potential energy, is not conserved and decreases with time. So,
one might think that (\ref{e1}) cannot be derived from a time-independent
Hamiltonian.

However, Bateman \cite{r1} made the remarkable observation that if one appends
the time-reversed oscillator equation with undamping (gain) instead of damping,
\begin{equation}
\ddot{y}-2\gamma\dot{y}+\w^2y=0\qquad(\gamma>0),
\label{e3}
\end{equation}
then even though the two oscillators are independent and noninteracting, the two
equations of motion (\ref{e1}) and (\ref{e3}) can be derived from the
time-independent quadratic Hamiltonian
\begin{equation}
H=pq+\gamma(yq-xp)+\left(\w^2-\gamma^2\right)xy.
\label{e4}
\end{equation}
The two oscillator equations follow directly from Hamilton's equations of motion
\begin{eqnarray}
\dot{x}=\frac{\partial H}{\partial p}&=&q-\gamma x,\label{e5}\\
\dot{y}=\frac{\partial H}{\partial q}&=&p+\gamma y,\label{e6}\\
\dot{p}=-\frac{\partial H}{\partial x}&=&\gamma p-\left(\w^2-\gamma^2\right)
y,\label{e7}\\
\dot{q}=-\frac{\partial H}{\partial y}&=&-\gamma q-\left(\w^2-\gamma^2\right)x.
\label{e8}
\end{eqnarray}
To derive (\ref{e1}) we differentiate (\ref{e5}) with respect to $t$, eliminate
$\dot{q}$ by using (\ref{e8}), and eliminate $q$ by using (\ref{e5}). Similarly,
to derive (\ref{e3}), we differentiate (\ref{e6}) with respect to $t$, eliminate
$\dot{p}$ by using (\ref{e7}), and eliminate $p$ by using (\ref{e6}).

The Hamiltonian (\ref{e4}) is $\cPT$ symmetric \cite{r2}; under parity
reflection $\cP$ the oscillators with loss and gain are interchanged,
\begin{equation}
\cP:\,x\to y,\quad y\to x,\quad p\to q,\quad q\to p,
\label{e9}
\end{equation}
and under time reversal $\cT$ the signs of the momenta are reversed,
\begin{equation}
\cT:\,x\to x,\quad y\to y,\quad p\to-p,\quad q\to-q.
\label{e10}
\end{equation}
[The Hamiltonian $H$ in (\ref{e4}) is $\cPT$ symmetric but it is not invariant
under $\cP$ or $\cT$ separately.] Because the balanced-loss-gain system is
described by a time-independent Hamiltonian, the energy (the value of $H$) is
conserved in time. However, the energy has the complicated form in (\ref{e4})
and is not a simple sum of kinetic and potential energies. If the loss and gain
terms in (\ref{e1}) and (\ref{e3}) were not exactly balanced (that is, if the
velocity terms did not have equal but opposite signs), the system would not be
derivable from a time-independent quadratic Hamiltonian.

It is even more remarkable that the equation of motion (\ref{e1}) of the damped 
oscillator can be derived from a (nonquadratic) time-independent Hamiltonian
{\it without having to introduce any additional degrees of freedom}. The
construction of such a Hamiltonian was first given in Ref.~\cite{r3} and was
accomplished by using a rather obscure technique called the {\it Prelle-Singer
method}.

In this paper we show how to construct the Hamiltonian for an arbitrary
homogeneous linear constant-coefficient differential equation of {\it any}
order. First, in Sec.~\ref{s2} we do so for the second-order equation (\ref{e1})
and in Sec.~\ref{s3} we demonstrate the procedure for a general third-order
equation. An interesting special case of such an equation is the equation that
describes the nonrelativistic self-acceleration of a charged oscillating
particle \cite{r4} and it is quite remarkable that even though there are
runaway modes, the energy of such a system is conserved. Then, in Sec.~\ref{s4}
we generalize our procedure to an arbitrary $n$th-order constant-coefficient
equation. Section~\ref{s5} discusses the problem of quantization and we show
that quantizing the classical Hamiltonians discussed in this paper is
nontrivial. Finally, Sec.~\ref{s6} gives a brief summary.

\section{Hamiltonian for a general linear constant-coefficient second-order
differential equation}
\label{s2}
We begin with (\ref{e1}) and substitute $x(t)=e^{-i\nu t}$. This gives a
quadratic equation for the frequency $\nu$:
\begin{equation}
\nu^2+2i\gamma\nu-\w^2=0.
\label{e11}
\end{equation}
This equation factors
\begin{equation}
\left(\nu-\w_1\right)\left(\nu-\w_2\right)=0,
\label{e12}
\end{equation}
where 
\begin{equation}
\w_1+\w_2=-2i\gamma,\qquad\w_1\w_2=-\w^2,
\label{e13}
\end{equation}
and thus
\begin{equation}
\w_{1,2}=-i\gamma\pm\Omega=-i\gamma\pm\sqrt{\w^2-\gamma^2}.
\label{e14}
\end{equation}

We claim that a Hamiltonian $H$ that generates a general linear
constant-coefficient $n$th-order evolution equation has the generic form
\begin{equation}
H=axp+f(p),
\label{e15}
\end{equation}
where $a$ is a constant and $f(p)$ is a function of $p$ only. For the case of
the second-order equation (\ref{e1}), one such Hamiltonian is
\begin{equation}
H_1=-i\w_1xp+\frac{\w_1}{\w_1-\w_2}p^{1-\w_2/\w_1}.
\label{e16}
\end{equation}
A second and equally effective Hamiltonian is obtained by interchanging the
subscripts $1$ and $2$:
\begin{equation}
H_2=-i\w_2xp+\frac{\w_2}{\w_2-\w_1}p^{1-\w_1/\w_2}.
\label{e17}
\end{equation}
These Hamiltonians appear in Ref.~\cite{r3} for the case of over-damping
($\gamma^2>\w^2$), in which case they are real, but they apply equally well when
($\gamma^2<\w^2$). (We are not concerned here with the reality of the
Hamiltonian.)

For the Hamiltonian $H_1$, Hamilton's equations read
\begin{eqnarray}
\dot{x}&=&\frac{\partial H_1}{\partial p}=-i\w_1x+p^{-\w_2/\w_1},
\label{e18}\\
\dot{p}&=&-\frac{\partial H_1}{\partial x}=i\w_1p.\label{e19}
\end{eqnarray}
We then take a time derivative of (\ref{e18}) and simplify the resulting
equation first by using (\ref{e19}) and then by using (\ref{e18}):
\begin{eqnarray}
\ddot{x}+i\w_1\dot{x}&=&-\frac{\w_2}{\w_1}p^{-1-\w_2/\w_1}
\dot{p}\nonumber\\
&=&-i\w_2p^{-\w_2/\w_1}\nonumber\\
&=&-i\w_2\left(\dot{x}+i\w_1x\right).
\label{e20}
\end{eqnarray}
Thus, 
\begin{equation}
\ddot{x}+i\left(\w_1+\w_2\right)\dot{x}-\w_1\w_2x=0,
\label{e21}
\end{equation}
which reduces to (\ref{e1}) upon using (\ref{e13}).

The evolution equation (\ref{e1}) has one conserved (time-independent) quantity,
and this quantity can be expressed in terms of the function $x(t)$ only. To find
this quantity, we begin with (\ref{e18}) and solve for $p$:
\begin{equation}
p=\left(\dot{x}+i\w_1x\right)^{-\w_1/\w_2}.
\label{e22}
\end{equation}
We then use this result to eliminate $p$ from the Hamiltonian $H_1$. Since
$H_1$ is time-independent, we conclude that 
\begin{equation}
C_1=\frac{\left(\dot{x}+i\w_2x\right)^{\w_2}}{\left(\dot{x}+i\w_1x\right)^{
\w_1}}
\label{e23}
\end{equation}
is conserved. Had we started with the Hamiltonian $H_2$ we would have obtained
the conserved quantity
\begin{equation}
C_2=\frac{\left(\dot{x}+i\w_1x\right)^{\w_1}}{\left(\dot{x}+i\w_2x\right)^{
\w_2}},
\label{e24}
\end{equation}
but this is not an independent conserved quantity because $C_2=1/C_1$. These
conserved quantities were also found in Ref.~\cite{r3} for the case of
over-damping.

When $\gamma=0$, these results reduce to the familiar expressions in the case of
the simple harmonic oscillator. In this case we let $\w=\w_1=-\w_2$ so that
$H_1$ becomes 
\begin{equation}
H_1=-i\w xp+\half p^2,
\label{e25}
\end{equation}
which is related to the standard simple harmonic oscillator Hamiltonian by the 
change of variable $p\to p-i\w x$. The conserved quantities $C_2$ and $C_1$
become simply $(\dot{x}^2+\w^2x^2)^{\pm\w}$, in which we recognize the usual
conserved total energy.

\section{Hamiltonian for a constant-coefficient third-order equation}
\label{s3}

In this section we show how to construct a Hamiltonian that gives rise to the
general third-order constant-coefficient evolution equation
\begin{equation}
(D+i\w_1)(D+i\w_2)(D+i\w_3)x=0,
\label{e26}
\end{equation}
where $D\equiv\frac{d}{dt}$. The Hamiltonian that we will construct has just one
degree of freedom.

An interesting physical example of such a differential equation is the
third-order differential equation
\begin{equation}
m\ddot{x}+kx-m\tau\dddot{x}=0
\label{e27}
\end{equation}
that describes an oscillating charged particle subject to a radiative
back-reaction force \cite{r4}. Following Bateman's approach for the damped
harmonic oscillator, Englert \cite{r5} showed that the pair of noninteracting
equations (\ref{e27}) and 
\begin{equation}
m\ddot{y}+ky+m\tau\dddot{y}=0
\label{e28}
\end{equation}
can be derived from the quadratic Hamiltonian
\begin{equation}
H=\frac{ps-rq}{m\tau}+\frac{2rs}{m\tau^2}+\frac{pz+qw}{2}-\frac{mzw}{2}+kxy.
\label{e29}
\end{equation}
This Hamiltonian contains the {\it four} degrees of freedom $(x,p)$, $(y,q)$,
$(z,r)$, and $(w,s)$. An interacting version of this model was studied in
Ref.~\cite{r6}. In fact, we find that the two equations of motion (\ref{e27})
and (\ref{e28}) can be derived from the simpler quadratic Hamiltonian
$$H=\frac{pr+qz}{\sqrt{m\tau}}-\frac{rz}{\tau}+kxy,$$
which has only the three degrees of freedom $(x,p)$, $(y,q)$, and $(z,r)$. A
similar three-degree of freedom Hamiltonian was also found in Ref.~\cite{r5}.

Our objective here is to find a {\it one}-degree-of-freedom Hamiltonian that can
be used to derive the third-order differential equation (\ref{e26}). Note that
the general solution to (\ref{e26}) is
\begin{equation}
x=a_1e^{-i\w_1t}+a_2 e^{-i\w_2t}+a_3 e^{-i\w_3t},
\label{e30}
\end{equation}
where $a_k$ are arbitrary constants. If we form $(D+i\w_2)(D+i\w_3)x$, that is,
$\ddot{x}+i(\w_2+\w_3)\dot{x}-\w_2\w_3x$, we obtain
\begin{equation}
a_1e^{-i\w_1t}=-\frac{(D+i\w_2)(D+i\w_3)x}{(\w_1-\w_2)(\w_1-\w_3)}
\label{e31}
\end{equation}
in which the constants $a_2$ and $a_3$ do not appear. Similarly, we have
\begin{eqnarray}
a_2e^{-i\w_2t}&=&-\frac{(D+i\w_3)(D+i\w_1)x}{(\w_2-\w_3)(\w_2-\w_1)},\nonumber\\
a_3e^{-i\w_3t}&=&-\frac{(D+i\w_1)(D+i\w_2)x}{(\w_3-\w_1)(\w_3-\w_2)}.
\label{e32}
\end{eqnarray}
So, assuming that the frequencies $\w_k$ are all distinct, there are two
independent conserved quantities, namely
\begin{eqnarray}
C_2=\frac{[\ddot{x}+i(\w_1+\w_2)\dot{x}-\w_1\w_2 x]^{1/\w_3}}{[\ddot{x}+i
(\w_2+\w_3)\dot{x}-\w_2\w_3 x]^{1/\w_1}},\nonumber\\
C_3=\frac{[\ddot{x}+i(\w_1+\w_3)\dot{x}-\w_1\w_3 x]^{1/\w_2}}{[\ddot{x}+i
(\w_2+\w_3)\dot{x}-\w_2\w_3 x]^{1/\w_1}}. 
\label{e33}
\end{eqnarray}

These expressions and the equation of motion can be derived from the Hamiltonian
\begin{equation}
H=-i\w_1 xp+\frac{b_2\w_1}{\w_1-\w_2} p^{1-\w_2/\w_1}+\frac{b_3\w_1}{\w_1-\w_3}
p^{1-\w_3/\w_1},
\label{e34}
\end{equation}
where $b_2$ and $b_3$ are arbitrary constants.
Thus, $\dot{p}\equiv-\frac{\partial H}{\partial x}=i\w_1 p$. This means that $p
\propto e^{i\w_1 t}$, so that $1/p$ is directly related to the combination
in (\ref{e31}).

Then, from Hamilton's equation $\dot{x}\equiv\frac{\partial H}{\partial p}$ and
from further differentiation with respect to $t$, we obtain
\begin{eqnarray}
\dot{x}&=&-i\w_1 x+b_2 p^{-\w_2/\w_1}+b_3 p^{-\w_3/\w_1},\nonumber\\
\ddot{x}&=&-i\w_1 \dot{x}-i\w_2 b_2 p^{-\w_2/\w_1}-i\w_3 b_3 p^{-\w_3/\w_1},\\ 
\nonumber
\dddot{x}&=&-i\w_1 \ddot{x}-\w_2^2 b_2 p^{-\w_2/\w_1}-\w_3^2 b_3 p^{-\w_3/\w_1}.
\label{e35}
\end{eqnarray}
These equations depend on the constants $b_2$ and $b_3$. Nevertheless, after we
combine these equations and perform some simplifying algebra, we obtain
\begin{equation}
\dddot{x}+i(\w_1+\w_2+\w_3)\ddot{x}-(\w_1\w_2+\w_2\w_3+\w_3\w_1)\dot{x}-i
\w_1\w_2\w_3x=0.
\label{e36}
\end{equation}
The constants $b_2$ and $b_3$ have disappeared in this combination and we have
reconstructed the equation of motion (\ref{e26}).

Using only derivatives up to the second order, we can find expressions for $b_2
p^{-\w_2/\w_1}$ and $b_3p^{-\w_3/\w_1}$, namely
\begin{eqnarray}
i(\w_3-\w_2)b_2p^{-\w_2/\w_1}&=&\ddot{x}+i(\w_1+\w_3)\dot{x}-\w_1\w_3x,
\nonumber\\
i(\w_2-\w_3)b_3 p^{-\w_3/\w_1}&=& \ddot{x}+i(\w_1+\w_2)\dot{x}-\w_1\w_2 x.
\label{e37}
\end{eqnarray}
These are precisely the combinations appearing in (\ref{e33}), and from them we
can construct the conserved quantity $C_2/C_3$.

In order to derive the second conserved quantity we use the fact that the
Hamiltonian is a constant. We then evaluate $H$ in terms of $x$, $\dot{x}$, and
$\ddot{x}$ using (\ref{e37}) and after some algebra we find that
\begin{equation}
H=i\w_1p\,\frac{\ddot{x}+i(\w_2+\w_3)\dot{x}-\w_2\w_3x}{(\w_1-\w_2)(\w_1-\w_3)},
\label{e38}
\end{equation}
which contains precisely the other linear combination of $\ddot{x}$, $\dot{x}$,
and $x$ that appeared in (\ref{e31}). As already mentioned, $1/p$ is
proportional to that combination.

To summarize, the evolution equation (\ref{e26}) can be derived from the unusual
time-independent Hamiltonian (\ref{e34}) containing the single coordinate
variable $x$ and its conjugate momentum $p$. This Hamiltonian is a conserved
quantity, which can be expressed as a function of $\ddot{x}$, $\dot{x}$, and
$x$. There is a second conserved quantity having a similar structure. In the
next section we show how this procedure generalizes to a linear
constant-coefficient differential equation of order $n$, for which there are
$n-1$ conserved quantities involving time derivatives up to order $n-1$.

However, before moving on, it is important to resolve an apparent paradox,
namely, how a Hamiltonian with a single degree of freedom can give rise to a
differential equation whose order is greater than two. The problem is this: Our
Hamiltonian has the general form in (\ref{e15}): $H=axp+f(p)$. Therefore, the
equations of motion are simply
\begin{equation}
{\dot x}=ax+g(p),
\label{e39}
\end{equation}
where $g(p)=f'(p)$, and
\begin{equation}
{\dot p}=-ap
\label{e40}
\end{equation}
We solve (\ref{e40}) first,
\begin{equation}
p(t)=Ce^{-at},
\label{e41}
\end{equation}
where C is an arbitrary constant. Next, we return to (\ref{e39}), which becomes
\begin{equation}
{\dot x}=ax+g\left(Ce^{-at}\right)
\label{e42}
\end{equation}
after we eliminate $p$ by using (\ref{e41}). This is a {\it first-order}
equation. Thus, its solution has only {\it two} arbitrary constants:
\begin{equation}
x(t)=\phi(t,C,D).
\label{e43}
\end{equation}
We obtained the higher-order differential equation (\ref{e30}) by the sequence
of differentiations in (\ref{e36}). However, the solution to an $n$th-order
equation can incorporate $n$ pieces of data such as $n$ initial conditions:
$x(0)$, ${\dot x}(0)$, ${\ddot x}(0)$, ${\dddot x}(0)$, and so on. How is it
possible to incorporate $n$ pieces of data with only two arbitrary constants $C$
and $D$? There appear to be $n-2$ missing arbitrary constants.

The surprising answer is the $n-2$ pieces of initial data determine $n-2$
parameters multiplying each of the fractional powers of $p$ in $H$. (One
parameter can always be removed by a scaling.) We call these parameters $b_k$.
Thus, we have incorporated the initial data into the Hamiltonian in the form of
{\it coupling-constant} parameters.

For the triple-dot equation, we can see from (\ref{e37}) that the ratio $b_2^{1/
\w_2}/b_1^{1/\w_1}$ is related to the initial conditions. So, for the case of
the third-order equation, the three arbitrary constants are $C$, $D$, and
$b_2^{1/\w_2}/b_1^{1/\w_1}$. We emphasize that if one just wants a Hamiltonian
that gives the equations of motion, the coefficients $b_k$, which play the role
of coupling constants, are irrelevant and we have shown that they drop out from
the equation of motion. However, the coupling constants in the Hamiltonian are
required to incorporate the initial data and are determined by the initial data.

\section{Hamiltonian for a constant-coefficient $n$th-order equation}
\label{s4}

It is straightforward to generalize to the case of an arbitrary $n$th-order
constant-coefficient evolution equation
\begin{equation}
\left[\prod_{r=1}^n\left(D+i\w_r\right)\right]x(t)=0,
\label{e44}
\end{equation}
whose general solution is
\begin{equation}
x(t)=\sum_{r=1}^n a_re^{-i\w_rt}.
\label{e45}
\end{equation}
For simplicity, we assume first that the frequencies $\w_r$ are all distinct;
at the end of this section we explain what happens if some of the
frequencies are degenerate.

Corresponding to (\ref{e32}) and (\ref{e33}), we have
\begin{equation}
e^{-i\w_st}\propto\left[\prod_{r\neq s}\left(D+i\w_r\right)\right]x(t).
\label{e46}
\end{equation}
Thus, the quantity
\begin{equation}
Q_s\equiv\left\{\left[\prod_{r\neq s}^n\left(D+i\w_r\right)\right]x(t)
\right\}^{1/\w_s}
\label{e47}
\end{equation}
is proportional to $e^{-it}$ for all $s$.
Hence, the $n-1$ independent ratios $Q_s/Q_1$ ($s>1$) are all conserved. Any
other conserved quantities can be expressed in terms of these ratios.

The equation of motion and the conserved quantities
can be derived from the Hamiltonian
\begin{equation}
H=-i\w_1 xp+\sum_{r\neq1}^n \frac{b_r\w_1 p^{1-\w_r/\w_1}}{\w_1-\w_r},
\label{e48}
\end{equation}
which is the $n$th order generalization of (\ref{e34}) for the cubic case. In
this expression the $n-1$ coefficients $b_r$ are arbitrary. Note that in
constructing the Hamiltonian $H$ there is nothing special about the subscript
``1'' and it may be replaced by the subscript ``$s$'' ($1<s\leq n$).

\vspace{.2cm}
\leftline{\bf{Degenerate frequencies}}
Until now, we have assumed that the frequencies $\w_r$ are all distinct.
However, if some of the frequencies are degenerate, there is a simple way to
construct the appropriate Hamiltonian: If the frequencies $\w_1$ and $\w_2$
are equal, we make the replacement
\begin{equation}
\frac{\w_1}{\w_1-\w_2} p^{1-\w_2/\w_1}\longrightarrow\log(p).
\label{e49}
\end{equation}
(In making this replacement we are shifting the Hamiltonian by an infinite
constant.) Thus, for $\w_1=\w_2$ the Hamiltonian $H_1$ in (\ref{e16}) reduces to
\begin{equation}
H_1=-i\w_1 xp+\log p.
\label{e50}
\end{equation}
Hamilton's equations for this Hamiltonian immediately simplify to
(\ref{e21}) with $\w_1=\w_2$.

Similarly, for the case $\w_1=\w_2$ the Hamiltonian (\ref{e34}) reduces to
\begin{equation}
H=-i\w_1 xp+b_2\log p+\frac{b_3\w_1}{\w_1-\w_3}p^{1-\w_3/\w_1}
\label{e51}
\end{equation}
and Hamilton's equations for this Hamiltonian readily simplify to (\ref{e36})
with $\w_1=\w_2$.

Also, if the frequencies are triply degenerate, $\w_1=\w_2=\w_3=\w$, the
Hamiltonian in (\ref{e34}) is replaced by
\begin{equation}
H=-i\w xp+b\log p+\half c(\log p)^2,
\label{e52}
\end{equation}
where $b$ and $c$ are two parameters that are determined by the initial data.
Once again, Hamilton's equations for this Hamiltonian combine to give
(\ref{e36}) with $\w_1=\w_2=\w_3=\w$.

\section{Quantization}
\label{s5}
The obvious question to be addressed next is whether it is possible to use the
Hamiltonians that we have constructed to quantize classical systems that obey
linear constant-coefficient evolution equations. Let us begin by discussing the
simple case of the quantum harmonic oscillator (QHO), whose Hamiltonian $H_1$
is given in (\ref{e25}).

One possibility is to quantize Hamiltonian in $p$-space by setting $x=id/dp$.
Then the time-independent Schr\"odinger eigenvalue equation is, by shifting $E$
by $\w$
\begin{equation}
H_1 \tilde{\psi}(p)=\left(\w p\frac{d}{dp}+\half p^2\right)\tilde{\psi}(p)=
E\tilde{\psi}(p),
\label{e53}
\end{equation}
whose solution is
\begin{equation}
\tilde{\psi}(p)\propto p^{E/\w}e^{-p^2/(4\w)^2}.
\label{e54}
\end{equation}
In this way of doing things we can derive the quantization condition by
demanding that $\tilde{\psi}$ be a well-defined, nonsingular function, which
requires that $E=n\w$, where $n$ is a nonnegative integer \cite{r7}. However,
these ``momentum-space" eigenfunctions are problematical because $p$ has no
clear physical interpretation as a momentum, and it is not a Hermitian operator.
The momentum eigenfunctions are certainly not orthonormal in any simple sense
because they do not solve a Sturm-Liouville boundary-value problem \cite{r8}.

However, we can calculate the corresponding $x$-space eigenfunctions by Fourier
transform using the formula \cite{r9}
\begin{equation}
H_n(z)=\frac{(-i)^n}{2\sqrt{\pi}}e^{z^2}\int_{-\infty}^\infty dp\ e^{ipz} p^n
e^{-p^2}.
\label{e55}
\end{equation}
We find that
\begin{equation}
\psi_n(x)\propto e^{-\half \w^2 x^2 }\varphi_n(x),
\label{e56}
\end{equation}
where $\varphi_n(x)$ is the $n$th eigenfunction of the QHO. This is consistent
with our remark above that $H_1$ is related to the standard QHO Hamiltonian by
the transformation $p\to p-i\w x$. This transformation is achieved at the
operator level by the similarity transformation $p\to e^{-\w^2x^2/2}p e^{\w^2
x^2/2}$ \cite{r10}. Because of this additional factor, our eigenfunctions are
orthonormal with respect to the metric $\eta=e^{\w^2 x^2}$. As an alternative
approach, we can cast (\ref{e25}) in $x$-space as
\begin{equation}
H_1=-\half\frac{d^2}{dx^2}-\w^2\left(1+x\frac{d}{dx}\right),
\label{e57}
\end{equation}
from which we can obtain the $\psi_n(x)$ directly.


To summarize, the quantized version of (\ref{e25}) corresponds to a transformed
version of the QHO, where the $x$-space eigenfunctions are simply related to the
standard eigenfunctions, and are orthonormal with respect to an additional
weight function. The $p$-space eigenfunctions can be written down but their
interpretation is not at all obvious (the operator $p$ corresponds to the
conventional raising operator $a^\dag$) and are not orthogonal in any simple
way. In $p$ space the weight function $e^{\w^2 x^2}$ becomes the highly nonlocal
operator $e^{-\w^2 d^2/dp^2}$.

If we now generalize to the damped harmonic oscillator, we can still find a
solution $\tilde\psi(p)$ to the time-independent Schr\"odinger equation, namely
\begin{equation}
\tilde{\psi}(p)\propto p^{E/\w_1}\exp\left[-\frac{\w_1}{(\w_1-\w_2)^2}p^{1-
\w_1/w_2}\right],
\label{e58}
\end{equation}
but even if we take $E=n\w_1$ in order to make the prefactor nonsingular, we are
still left with a nonintegral, and in general complex, power of $p$ in the
exponential. (See, also, the comments in Ref.~\cite{r3}.) Thus, in addition to
the previously discussed problems with $\tilde {\psi}(p)$, we would now have to
consider it to be a function in a cut plane. Moreover, there is no simple
formula like (\ref{e55}) whereby one could obtain the $x$-space eigenfunctions.
Furthermore, if we cast the equation in $x$-space we obtain
\begin{equation}
H_1=\frac{1}{1-\w_2/\w_1}\left\{-i\frac{d}{dx}\left[\left(-i\frac{d}{dx}
\right)^{-\w_2/\w_1}-i(\w_1-\w_2)x\right]\right\},
\label{e59}
\end{equation}
in which the difficulty associated with a fractional derivative is manifest.

We conclude that quantizing Hamiltonians of the form in (\ref{e15}) is
nontrivial. The problem of quantizing the cubic equation describing the
back-reaction force on a charged particle was solved in Ref.~\cite{r6}. However,
the system that was actually quantized was a {\it pair} of {\it coupled} cubic
equations in the {\it unbroken} $\cPT$-symmetric region. Thus, it may be that
the most effective approach for quantizing a Hamiltonian of the form (\ref{e15})
is to introduce a large number of additional degrees of freedom.

\section{Summary}
\label{s6}

We have shown that any $n$th-order constant-coefficient evolution equation can
be derived from a simple Hamiltonian of the form (\ref{e15}). Remarkably, this
Hamiltonian only has one degree of freedom, that is, one pair of dynamical
variables $(x,p)$. Furthermore, we have shown that for such a system there are
$n-1$ independent constants of the motion and we have constructed these
conserved quantities in terms of $x(t)$. However, we find that it is not easy to
formulate a general procedure to quantize the system described by the
Hamiltonian, and this remains an interesting open problem.

\end{document}